# Size-Resolved Photoelectron Anisotropy of Gas Phase Water Clusters and Predictions for Liquid Water

*Sebastian Hartweg[1], Bruce L. Yoder[1], Gustavo A. Garcia[2], Laurent Nahon[2], and Ruth Signorell[1,]\**

[1]Department of Chemistry and Applied Biosciences, Laboratory of Physical Chemistry, ETH Zürich, Vladimir-Prelog-Weg 2. CH-8093 Zürich, Switzerland

[2]Synchrotron SOLEIL, L'Orme des Merisiers, St. Aubin BP 48, 91192, Gif sur Yvette, France

*Correspondence to: rsignorell@ethz.ch.

**Date:** 12.12. 2016

**ABSTRACT**

We report the first measurement of size-resolved photoelectron angular distributions for the valence orbitals of water clusters with up to 20 molecules. A systematic decrease of the photoelectron anisotropy is found for clusters with up to 5-6 molecules, and most remarkably, convergence of the anisotropy for larger clusters. We suggest the latter to be the result of a local short-range scattering potential that is fully described by a unit of 5-6 molecules. The cluster data and a detailed electron scattering model are used to predict liquid water anisotropies. Reasonable agreement with experimental liquid jet data is found.

A detailed understanding of elastic and inelastic scattering of electrons in liquid water is of fundamental importance for the modelling of radiation damage in biological systems, the description of the behaviour of the solvated electron in chemistry, and for the quantitative interpretation of photoelectron spectra of liquid water and aqueous solutions [1-8]. For slow electrons (electron kinetic energy eKE $\lesssim$ 50 eV), detailed experimental scattering parameters (differential scattering cross sections and energy losses) were so far only reported for *amorphous ice* [9]; with the exception of very slow electrons (eKE $\lesssim$ 6 eV), for which *liquid water* data were recently obtained from photoelectron velocity map imaging (VMI) of liquid water droplets [10]. Since there is little reason to expect substantial differences between amorphous ice and liquid water for electronic scattering processes (eKEs $\gtrsim$ 6 eV) the amorphous ice and liquid droplet data [9,10] should now provide a reasonable data set for scattering simulations of liquid water. In addition, electron attenuation lengths (EALs) for eKEs $\gtrsim$ 3 eV are available for liquid water from various microjet studies [11-13], which, however, do not allow quantitative predictions of the scattering contributions.

The photoelectron angular distribution (PAD) is particularly sensitive to electron scattering and has thus recently received increasing attention in this context [7,10,13-18]. Often, the information in the PAD is described by a single anisotropy parameter *β* (see Eq. (1)). For the liquid microjet, this is an approximation which we also follow in the present work. For ionization from the O1s orbital of liquid water, Thürmer et al. observed a more isotropic PAD; i. e. a smaller *β*-value; for liquid water compared with gas phase water over the eKE range from ~12 - 450 eV [13]. For core-level ionization, this reduction is assumed to mainly arise from electron scattering within the liquid. For the ionization from the valence orbitals $1b_1$, $3a_1$, and $1b_2$, additional changes in the initial state due to orbital mixing also mediated by hydrogen-bonding are expected to contribute to the difference in *β*-values between gas and liquid phase. While monomer gas phase *β*-parameters have been reported for the three valence orbitals at photon energies 18 eV ≤ hv < 139 eV [15,16,19-21], corresponding values for liquid water have to the best of our knowledge only been reported at a single ionization energy of hv = 38.7 eV [16]. Zhang et al. [15] made a first attempt to distinguish between contributions to *β* arising from initial state effects versus those originating from electron scattering. This study is based on the measurement of $(H_2O)_n$ clusters with broad size distributions and estimated average sizes of $\langle n \rangle \geq 58$ at two ionization energies of hv = 40 and 60 eV. The results of ref. [15] point to the possibility of intrinsic differences between molecular and cluster PADs due to alterations in the initial states. The existing literature values for water cluster and liquid water *β*-parameters are summarized in Table T1 [22].

The present work reports double imaging photoelectron photoion coincidence measurements of small $(H_2O)_n$ clusters (n ≤ 20). As a unique feature this technique allows us to record photoelectron velocity map images (VMIs) for a particular cluster size n, and thus to extract cluster size-resolved *β*-parameters. This does not only avoid averaging of *β* over different cluster sizes, but it also prevents any issues from the overlap with the strong water monomer signal. Size selectivity is particularly important for small clusters, for which pronounced changes in *β* are expected for size changes by just one water molecule. Our main goal is to clarify the evolution of PADs as a function of cluster size. Clusters provide a link between the monomer and the liquid, and thus eventually contribute to a better understanding of the complex influence of electron scattering in liquid water. Towards this direction, we report calculated *β*-parameters for typical liquid water microjet experiments obtained with a detailed scattering model [7,9,10]. We focus on slow electrons with eKEs ≤ 65 eV; i. e. the range where the PADs sensitively depend on electron scattering.

VMIs of water clusters $(H_2O)_n$ (n = 1 - 20) were recorded with the double imaging photoelectron photoion coincidence ($i^2$PEPICO) spectrometer [23,24] available at the DESIRS VUV beamline [25] of the synchrotron radiation facility SOLEIL. For this project, the ion spectrometer provides a typical mass resolution of 1700 amu (FWHM), sufficient to separate the parent cluster ions $(H_2O)_n^+$ and the fragments $(H_2O)_{n-1}H^+$ (Fig. S1 [22] and Eq. (2)). The electron spectrometer yields eKE resolutions down to ~3 %. Coincident operation of the ion and electron analyzers allows the photoelectron images to be mass-tagged. The water clusters were produced by continuous supersonic expansions of water/helium gas mixtures (water pressure 0.2-1 bar, He pressure 3-7 bar) [26]. Images were recorded at twelve different photon energies *hv* between 12.5 and 35.0 eV with linearly polarized radiation and reconstructed with pBASEX [27] providing the radial and angular information we are interested in for a given cation mass. The normalized photoelectron angular distributions (PADs) are described by a single anisotropy parameter *β*, defined by

$$I(\theta) \propto 1 + \frac{\beta}{2}(3\cos^2\theta - 1) \quad \text{Eq. (1)}$$

The $\beta$-parameters for the gas phase monomer (n = 1) in the range 13.0 eV ≤ $hv$ ≤ 35.0 eV are shown in Figs. 1 and S2 [22] and the corresponding values are tabulated in Table T2 [22]. Fig. S2 [22] also provides a comparison with published data for $hv$ ≥ 18 eV [15,16,19,20]. Our monomer data are in good agreement with the literature values. In addition, we provide the first monomer data below ~18 eV, which clearly confirm the trend predicted by calculations [21] towards low anisotropies at very low photoelectron kinetic energies (eKEs between ~ 0.4 and 5.4 eV).

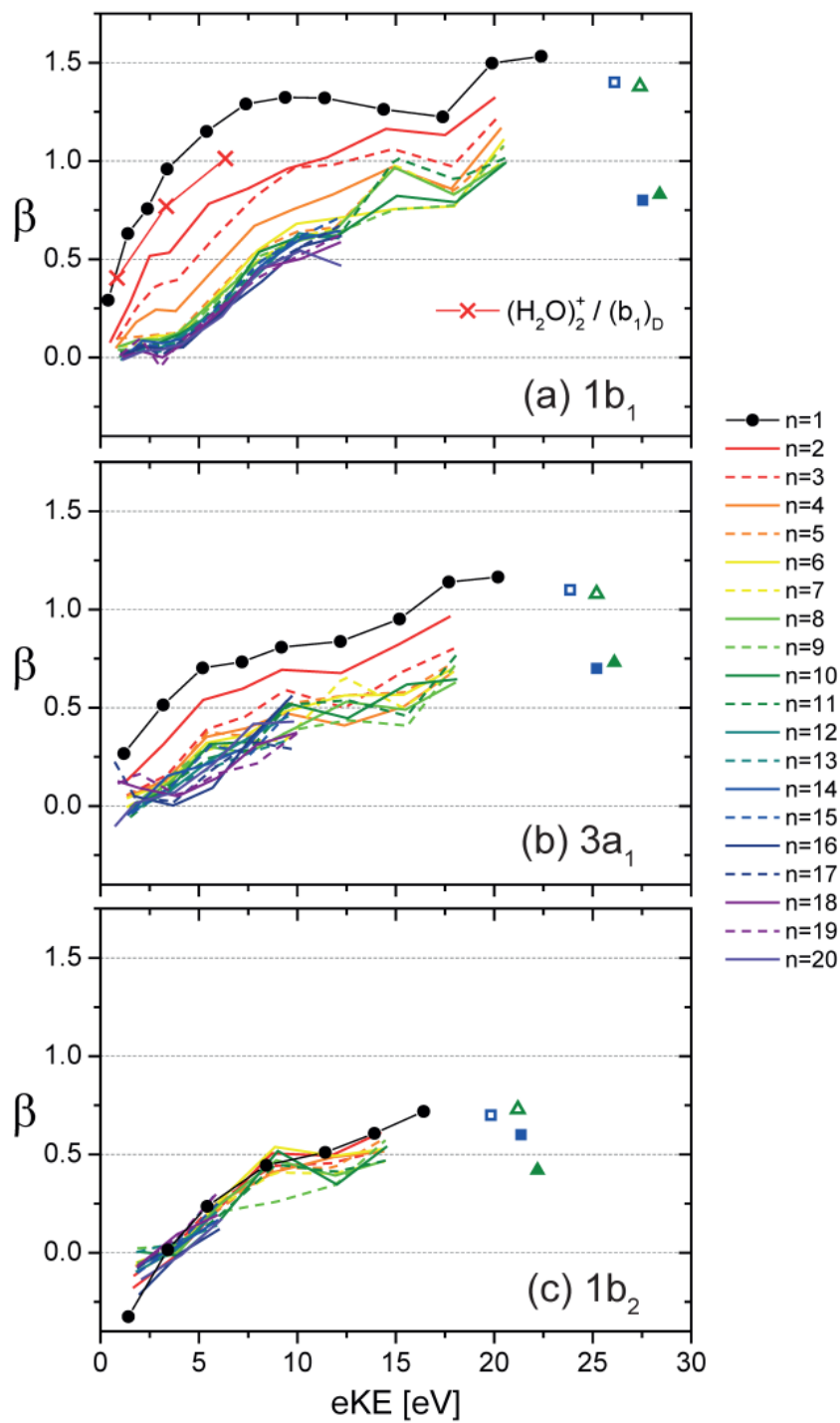

**Figure 1:** Lines: Size-resolved *β*-parameters for $(H_2O)_n$ clusters for n ≤ 20 recorded at 13.0 eV ≤ *hv* ≤ 35.0 eV. For $(H_2O)_2$, two *β*-traces are shown one for $H_3O^+$ (red full line; Eq. (2)) and one for the intact dimer $(H_2O)_2^+$ (red crosses, see Fig. 2). Green triangles: *β*-parameters for the monomer (open symbols) and a cluster ensemble (full symbols) with an average cluster size $\langle n \rangle \sim 58$ from ref. [15]. Blue squares: *β*-parameters for monomer (open symbols) and liquid water (full symbols) from the microjet study in ref. [16].

In addition to monomer data, Fig. 1 shows the summary of the experimental *β*-parameters for $(H_2O)_n$ cluster for n ≤ 20 recorded with i²PEPICO. The corresponding values with respective uncertainties are listed in Table T2 [22]. For larger clusters at higher *hv*, some data points are missing in Fig. 1 because the signal noise ratio was not sufficient to determine reliable *β*-parameters. Photoionization of a neutral water cluster $(H_2O)_n$ is accompanied by a fast intracluster proton transfer with subsequent loss of an OH radical [28-33]:

$$(H_2O)_n + hv = (H_2O)_{n-1} H^+ + OH + e^- \qquad \text{Eq. (2)}$$

According to Eq. (2), we assign clusters with n molecules to photoelectron images recorded in coincidence with cluster mass $m = (n \cdot 18) - 17$. Note, that for small clusters the subsequent slow loss of water molecules from the initially formed protonated cluster is dominated by monomer loss with total decay fractions of < 0.3 [28,29,34]. Exemplary photoelectron spectra and images for n = 1, 2 and 6 are shown in Figs. 2 and S3 [22], respectively. The vertical electron binding energy (VBE), i. e. the most probable electron binding energy (eBE), shifts to lower values for larger clusters due to polarization effects (Fig. S4 [22]), but the liquid bulk value [35] or the values of large clusters [15,36] are not yet reached. The downward trend in cluster VBEs is consistent with the evolution of the cluster ion appearance energy from ref. [28]. The dimer spectrum in Fig. 2 consists of the two contributions from the intact dimer $(H_2O)_2^+$ (red line) and from $H_3O^+$ (black line). In accordance with refs. [30,33], we assign the $(H_2O)_2^+$ contribution to arise primarily from the removal of an electron from the lone pair of the hydrogen-bond donor (referred to as $(b_1)_D$) and the $H_3O^+$ contribution to result from the ionization of an orbital that is delocalized over both hydrogen-bond donor and acceptor (referred to as $(a_1/b_1)$) leading to different *β*-parameters as clearly seen in Fig. 1. Table T3 [22] contains the corresponding VBEs of the dimer and a comparison with literature data.

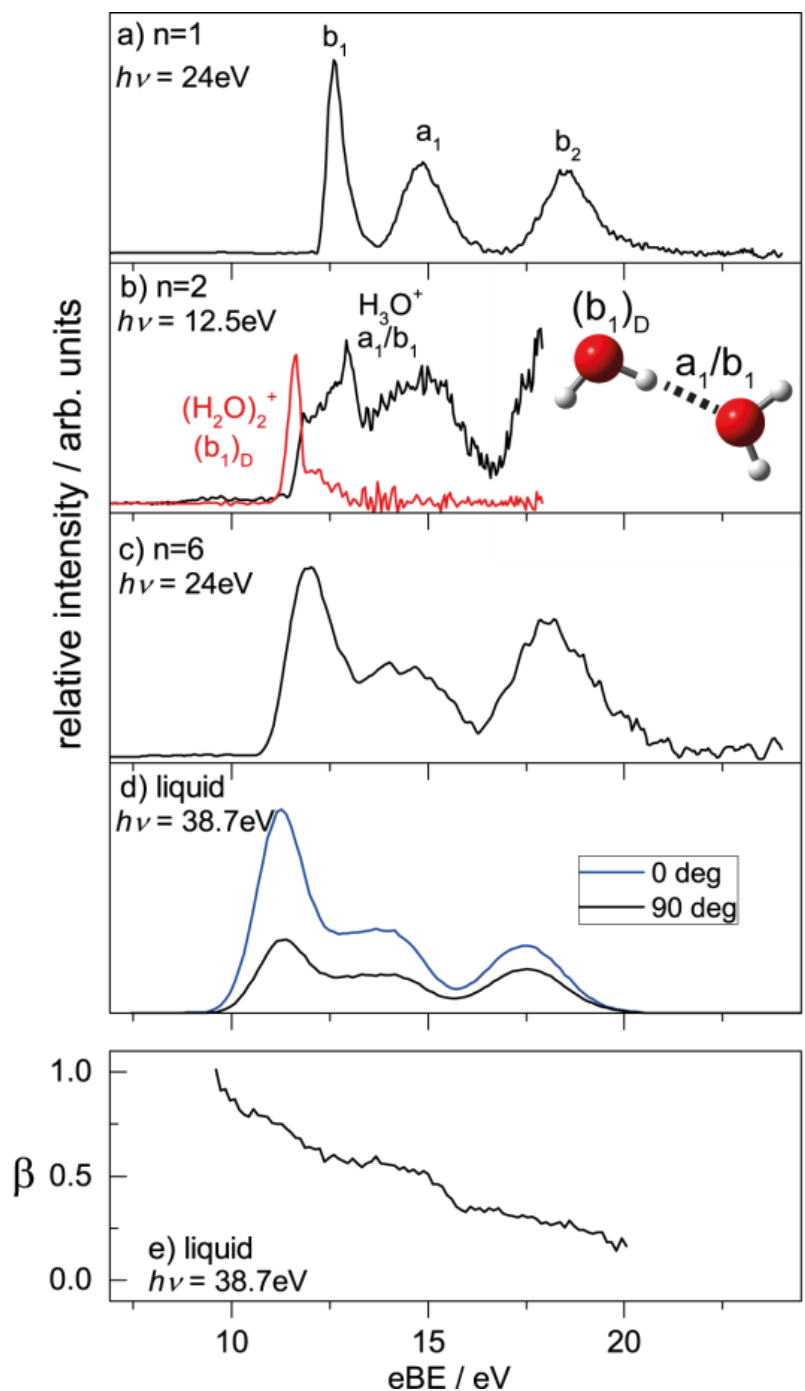

**Figure 2:** Experimental photoelectron spectra of **(a)** water monomer and $(H_2O)_n$ clusters for **(b)** n = 2 and **(c)** n = 6. Selected VMI images are shown in Fig. S3 [22]. The dimer spectrum has contributions from the intact dimer $(H_2O)_2^+$ (red line) and from $H_3O^+$ (black line) formed after fast proton transfer. **(d)** Calculated photoelectron spectra for the liquid water microjet for two polarization directions $\theta = 0°$ (blue line) and $90°$ (black line) of the light (Fig. S5 [22]) for monomer input $\beta_{n=1}$ (Fig. 3). **(e)** Calculated energy-dependent anisotropy parameter $\beta$ for liquid water. The photon energies $h\nu$ are indicated in the figure.

Fig. 1 provides the first quantitative $\beta$-values for the initial condensation steps. The largest absolute decrease in the photoelectron anisotropy with increasing cluster size is observed for the $1b_1$ orbital (out-of-plane lone pair), followed by a smaller decrease for the $3a_1$ orbital (in-plane lone pair). The $1b_2$ orbital ($\sigma_{OH}$ bonding orbital) shows the smallest variations in $\beta$, but because they lie within the estimated uncertainty no systematic trend with cluster size can be extracted from our data. Note that for simplicity we use here the monomer orbital nomenclature for the clusters, neglecting symmetry changes and orbital mixing. Qualitatively, a decrease in the sensitivity of $\beta$ from $1b_1$ to $1b_2$ upon condensation seems reasonable because of the strong influence of hydrogen-bonding on the $1b_1$ orbital. Similar trends compared with monomer data were observed for the cluster ensemble data at 40 eV [15] and the liquid microjet data at 38.7 eV [16]. The most striking result in Fig. 1 is the convergence of $\beta$ for the two outermost valence orbitals for cluster sizes with n ≳ 5-6. It is important to note here that slow cluster evaporation cannot be the origin of the observed convergence. This follows from a simple estimate based on reported total cluster decay fractions and the maximum number of monomers that can evaporate from a cluster after proton transfer [28,29,34].

We suggest the following qualitative explanation for the systematic decrease of $\beta$ with increasing n and the convergence of $\beta$ for n ≳ 5-6. The difference between molecular and cluster PAD arises from different contributions: (i) The first contribution comes from a change in the initial molecular electron wavefunction and thus from *a change in the orbital character* due to condensation. This includes polarization and orbital mixing. For $(H_2O)_n$ clusters, electron delocalization over hydrogen-bonds is likely a major factor here. For increasing cluster size, changes in the orbital character typically result in a decrease of $\beta$. (ii) The second contribution is again attributed to a change in the initial state, but this contribution is caused by *multicentre ionization*. The larger the cluster becomes the more equivalent units it has from which ionization can happen (quasi-degeneracy). Interference of partial waves from many centres leads to a decrease of $\beta$, which is thus likely to be more pronounced for larger cluster. (iii) The third contribution comes from a *change in the ion core potential*; i. e. the potential by which the outgoing electron wave is scattered. The biggest influence originates probably from the delocalization of the remaining positive charge through hydrogen-bonds. Again, this tends to lead to a decrease in $\beta$. Qualitatively, all three contributions favour more isotropic PADs; i. e. decreases in $\beta$, with increasing cluster sizes. Note that for larger clusters the observed $\beta$ is the average over several conformers. This expectation agrees with the experimental observation for cluster sizes up to n = 5-6 in Fig. 1. Furthermore, the observed convergence of $\beta$ for n ≳ 5-6 implies that the range of the contributions (i)-(iii) essentially extends over only a few molecules. n ≈ 5-6 coincides with the smallest cluster sizes for which three-dimensional hydrogen-bond networks become more stable than ring-topology structures; resulting in more than two hydrogen bonds per water molecule (refs. [37-40] and references therein). It is plausible that the typical range for changes in orbital character and in the ion core potential is approximately equal to the range of local hydrogen-bridges. Similarly interference effects due to multicentre ionization are also expected to be most pronounced just in a local environment. The convergence of $\beta$ for 6 ≲ n ≲ 20 agrees with an intrinsic, short-range scattering potential that is described by a cluster with n ≈ 6. Since the spatial extent of clusters with n ≲ 20 is very small (~ 7-10 Å) the long-range scattering potential is essentially an unshielded (vacuum) Coulomb-potential. Note that even semi-quantitative descriptions of the cluster PADs would require very high-level quantum chemical calculations [17,41-43], which are still a big challenge for such complex systems. Simple modelling approaches, such as gas phase scattering between the monomers in a cluster, are not suitable to describe the cluster behavior.

The water dimer is a special case because ionization from the lone pair of the hydrogen-bond donor $(b_1)_D$ is distinguishable from ionization of the mixed $(a_1/b_1)$ orbital, which is delocalized over donor and acceptor (Fig. 2b). The $\beta$-parameters for $(b_1)_D$ (red crosses in Fig. 1a) are slightly lower than the monomer value (n = 1, black circles). The $(b_1)_D$ orbital can be considered as a monomer orbital that is disturbed by the presence of the second $H_2O$ molecule. Since the $(b_1)_D$ orbital it is not directly involved in the hydrogen-bond the decrease in $\beta$ compared with the monomer is suggested to arise mainly from contribution (i). The even stronger decrease in the $\beta$ of the $(a_1/b_1)$ orbital (n = 2, full red line in Fig. 1a) is tentatively explained by the fact that here all three factors (i)-(iii) contribute. The trend in the two different dimer $\beta$-parameters is consistent with our above expectation that the contributions (i)-(iii) generally result in a decrease rather than an increase in $\beta$. However, it is important to note here that because of the nonlinear dependence of $\beta$-parameters on the angular photoelectron distribution absolute changes in $\beta$ are not a truthful measure of the magnitude of different contributions.

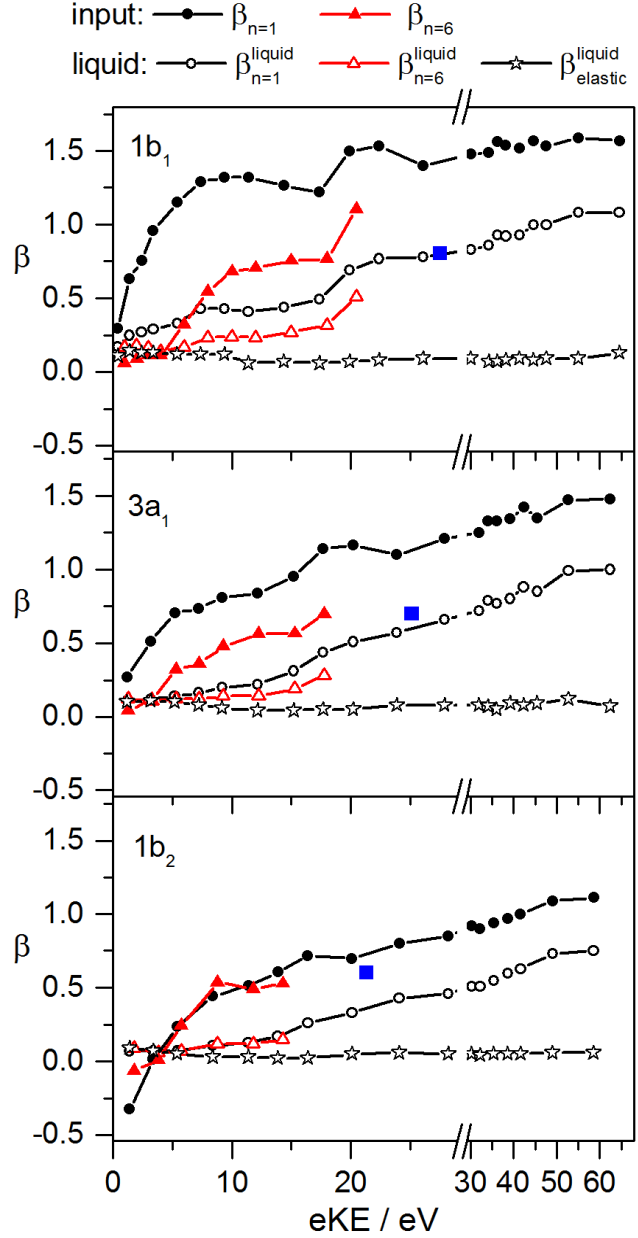

**Figure 3:** Calculated anisotropy parameters for liquid water microjets. Open black circles: $\beta_{n=1}^{liquid}$ calculated with the monomer values ($\beta_{n=1}$) as input for the local anisotropy in the liquid. Open red triangles: $\beta_{n=6}^{liquid}$ calculated with the cluster values ($\beta_{n=6}$) as input for the local anisotropy in the liquid. Open black stars: $\beta_{elastic}^{liquid}$ calculated with gas phase elastic scattering cross sections alone [44]. Full black circles: monomer values $\beta_{n=1}$ from this work (Fig. 1) and ref. [20]. Full red triangles: cluster values $\beta_{n=6}$ from this work (Fig. 1). Full blue squares: experimental $\beta^{liquid}$ by Faubel *et al.* [16].

The liquid water PAD is not only determined by the three local contributions (i)-(iii), but also by *elastic and inelastic scattering of the electrons within the liquid*. This fourth factor (iv) again results in a decrease in $\beta$. We simulate the contribution of this fourth factor (iv) to $\beta^{liquid}$ with a detailed scattering model [7,9,10] for a typical liquid microjet experiment, in which $\beta^{liquid}$ is determined from polarization-dependent measurements [12-14,16-18]. The scattering model and the retrieval of $\beta^{liquid}$ are described in section S2 in [22]. We assume the local contributions (i)-(iii) in liquid water to be either the same as in the monomer or as in a cluster with n = 6 (converged cluster value); i. e. we use either the experimental monomer ($\beta_{n=1}$) or the experimental cluster ($\beta_{n=6}$) anisotropy parameters from Fig. 1 to describe the local anisotropies in the liquid. We then calculate $\beta^{liquid}$ for the two different local input anisotropies with our scattering model. Calculated example photoelectron spectra for 0° and 90° laser polarization are shown in Fig. 2d together with the corresponding calculated $\beta^{liquid}$-values in Fig. 2e. The calculated photoelectron spectra agree well with experimental liquid-jet spectra [16,35]. Note that the spectra in ref. [16] contain large gas phase fractions. The resulting liquid anisotropy parameters from our scattering calculations, $\beta^{liquid}_{n=1}$ and $\beta^{liquid}_{n=6}$, respectively, are shown in Fig. 3. The comparison with the input values $\beta_{n=1}$ and $\beta_{n=6}$, respectively, shows the pronounced effect of contribution (iv) on the PADs. As expected it leads to a reduction of the anisotropy. Note that $\beta$-values retrieved from polarization-dependent liquid jet measurements are always marginally higher compared with $\beta$-values retrieved from other methods, e. g. VMI. This artifact is described in section S2 [22].

To the best of our knowledge experimental values for $\beta^{liquid}$ in the valence region were only reported at $hv$ = 38.7 eV from a microjet study by Faubel *et al.* [16] (Fig. 3, blue full squares). The agreement of our calculated $\beta^{liquid}$-values with the experimental data is reasonable. We conjecture that the larger deviations of the $a_1$ and $b_2$ values between experiment and simulation arise from overlapping monomer bands in the experiment, resulting in too high experimental liquid values. Compared with the experiment, we expect lower anisotropies from our model because it does not take into account the strong shielding of the ion core potential in the liquid. The range of the ion core potential is much reduced in the liquid compared with the monomer or the cluster case - an effect that is not represented by our current input values $\beta_{n=1}$ and $\beta_{n=6}$. Since better shielding means less scattering the inclusion of shielding presumably corresponds to higher input $\beta$-values and thus higher predicted $\beta^{liquid}$ -values. The apparent better agreement of the experimental data for the monomer input compared with the cluster input might be accidental. We expect that with the inclusion of shielding, the cluster input $\beta_{n=6}$ should provide better agreement with experimental liquid bulk data compared with the monomer input simply because the cluster input should better represent the local effects. A simple estimate of the influence of the shielding is unfortunately not possible. Again, such estimates require high-level ab initio calculations. We also add a calculation for the liquid anisotropy $\beta_{elastic}^{liquid}$ in Fig. 3, for which we used just *elastic gas phase* monomer scattering cross sections [44] instead of the proper condensed phase values as for the other simulations [9,10]. The resulting $\beta_{elastic}^{liquid}$ are essentially isotropic and clearly disagree with the experimental values at $h\nu$ = 38.7 eV. This demonstrates that gas phase scattering parameters are not suitable to describe the liquid.

Finally, we address the question to what extent contribution (iv) (elastic and inelastic scattering due to electron transport) arises in larger clusters (Fig. 1). To this end, we simulate cluster VMIs for different cluster sizes with our scattering model and determine $\beta$ from Eq. (1) [10,22]. For clusters with less than n ≈ 50 molecules, the influence of contribution (iv) is almost negligible. A significant deviation of $\beta$ on the order of 0.1 due to the influence of (iv) is only found for clusters with more than n ≈ 100 molecules (Fig. S6 [22]); i. e. beyond cluster sizes studied here.

In summary, photoelectron photoion concidence imaging provides size-dependent photoelectron anisotropy parameters of $(H_2O)_n$ clusters for n ≤ 20. The experimental data suggests that intracluster electron scattering in the size range between ~ 6 and 20 molecules is mainly determined by the short range potential of a cluster unit consisting of 5-6 molecules. This coincides with the smallest cluster sizes for which three-dimensional hydrogen-bond networks become the most stable structures. It seems reasonable that the short range scattering potential in liquid water is largely determined by this smallest unit; i. e. approximately by the first solvation shell. However, in contrast to the clusters, the ion core potential is strongly shielded in the liquid. At present, no reasonable estimate of the influence of shielding can be provided, but it might be argued that its inclusion will lead to an increase in the anisotropy compared with the experimental cluster data. We suspect that the major difference between small clusters (n ≲ 100) and liquid arises from the additional elastic and inelastic electron scattering within the liquid. A detailed scattering simulation for the liquid starting from cluster anisotropies of the smallest unit confirms this presumption. Even with the shielding effect neglected, this model provides reasonable agreement with experimental liquid jet data. Our simulations predict that gas phase scattering parameters are not appropriate for electron scattering within the liquid. Further validation of the role of the smallest cluster unit and the shielding in the liquid awaits more experimental data from liquid jets and larger water clusters as well as in-depth theoretical studies.

**Acknowledgment:** We thank Dr. David Luckhaus for help with the calculations and David Stapfer and Markus Steger for technical support. We are grateful to the SOLEIL staff for smoothly running the facility under project 20160122, particularly to Jean-François Gil and Duçan Bozanic for their technical support on the SAPHIRS experimental chamber. Financial support was provided by the ETH Zürich and the Swiss National Science Foundation under project no. 200020_159205.

**Supporting Information for:**

**Size-Resolved Photoelectron Anisotropy of Gas Phase Water Clusters and Predictions for Liquid Water**

*Sebastian Hartweg[1], Bruce L. Yoder[1], Gustavo A. Garcia[2], Laurent Nahon[2], and Ruth Signorell[1],\**

[1]Department of Chemistry and Applied Biosciences, Laboratory of Physical Chemistry, ETH Zürich, Vladimir-Prelog-Weg 2. CH-8093 Zürich, Switzerland

[2]Synchrotron SOLEIL, L'Orme des Merisiers, St. Aubin BP 48, 91192, Gif sur Yvette, France

*rsignorell@ethz.ch

**S1. Additional Figures S1-S6**

**S2. Scattering calculations for liquid water and clusters**

**S3. Additional Tables T1-T3**

**S4: References**

Date: 12.12.2016

## S1. Additional Figures

**Figure S1:** A time-of-flight mass spectrum showing a typical water cluster distribution. The measurement shown here was recorded upon photoionization with 16 eV photons. The neutral cluster distribution was created via continuous supersonic expansion of ~10% (0.5 bar) water vapor in 5 bar of helium. The inset shows the low mass region, including peak assignments.

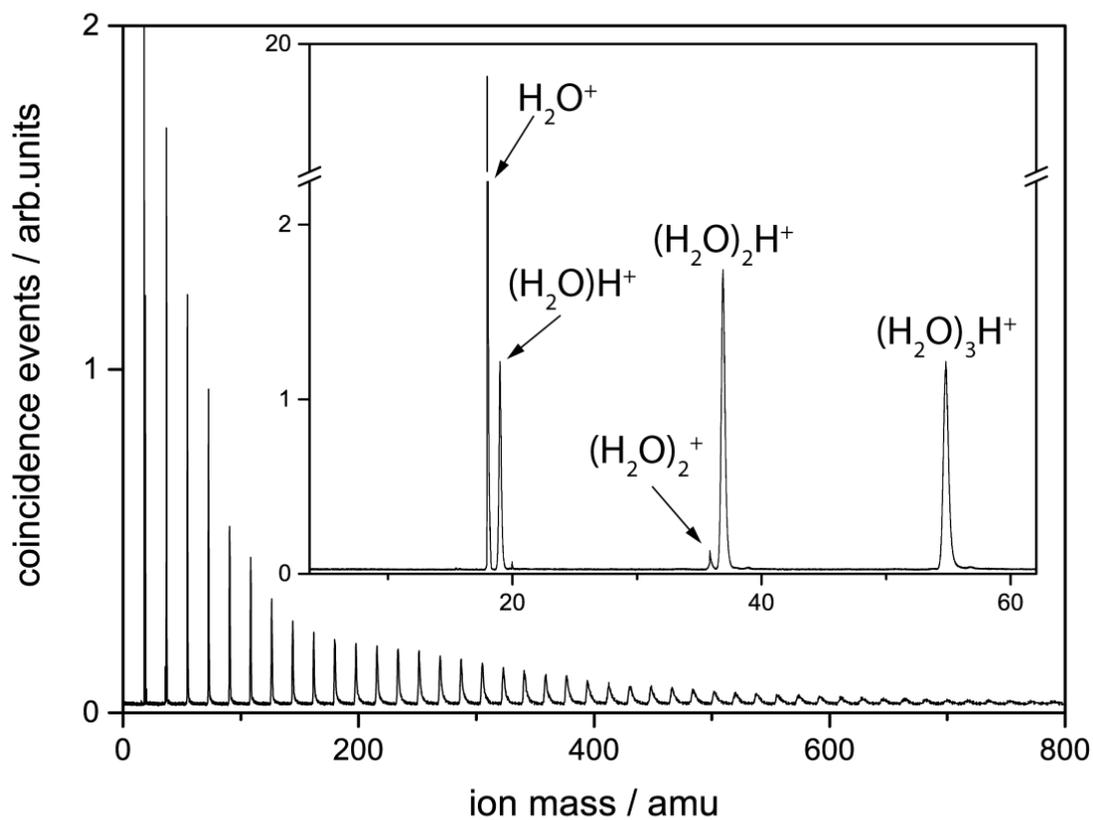

**Figure S2:** Experimental *β*-parameters of the gas phase monomer for the three valence orbitals **(a)** 1b$_1$, **(b)** 3a$_1$, and **(c)** 1b$_2$ for photon energies 13.0 eV $\leq h\nu <$ 65 eV. Full black circles: Present work; Open black diamonds: Truesdale et al. [1]; Red crosses: Banna et al. [2]; Open green triangles: Zhang et al. [3]; Open blue squares: Faubel et al. [4]. Our experimental monomer data confirm the trend towards low anisotropies at very low photoelectron kinetic energies (eKEs ~ 0.4 - 5.4 eV for 1b$_1$) predicted by calculations [5].

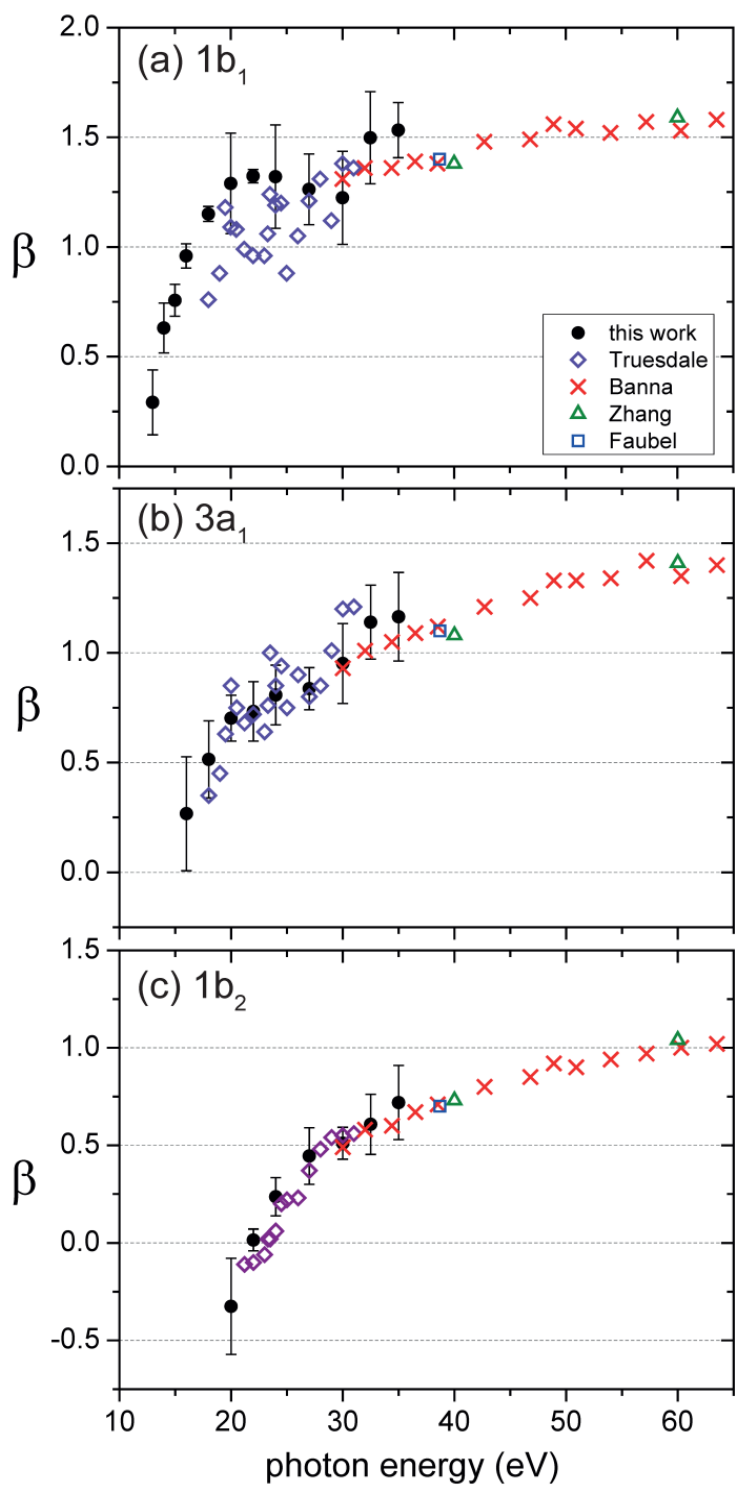

**Figure S3**: Reconstructed experimental velocity map photoelectron images recorded at $h\nu$ = 24 eV of **(a)** the gas phase monomer and **(b)** and the $(H_2O)_6$ water cluster. The pBasex program [6] was used for reconstruction. The three rings correspond to the valence orbitals $1b_1$ (outer ring), $3a_1$ (middle ring), and $1b_2$ (inner ring). The polarization and the propagation direction of the light are indicated by white arrows.

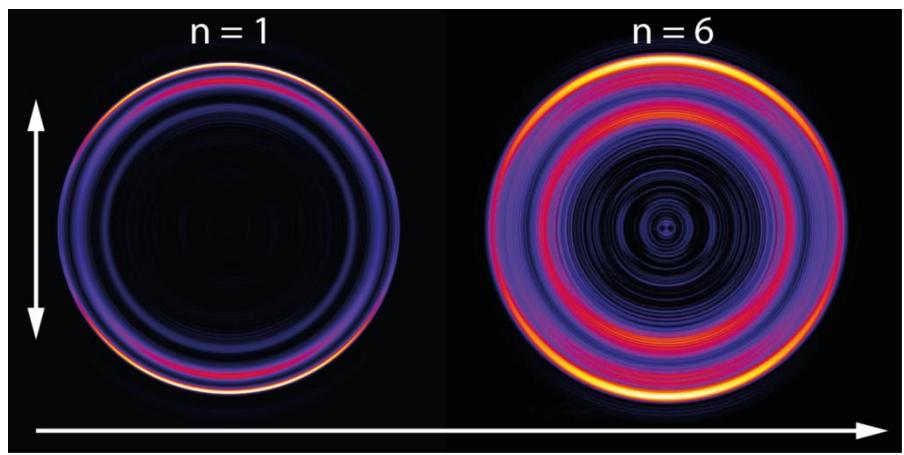

**Figure S4:** Band positions of **(a)** the $1b_1$, **(b)** the $3a_1$, and **(c)** the $1b_2$ band of $(H_2O)_n$ clusters as a function of cluster size n. The dimer values are not shown here, see Table T3. Crosses with thin lines: vertical binding energies (VBEs) from the present work. Open red circles: Cluster ion appearance energies (IAE) from ref. [7]. Full violet triangle: VBEs for a cluster ensemble with $\langle n \rangle > 160$ from ref. [8]. The dashed lines indicate the VBEs of liquid water from the liquid microjet study in ref. [9].

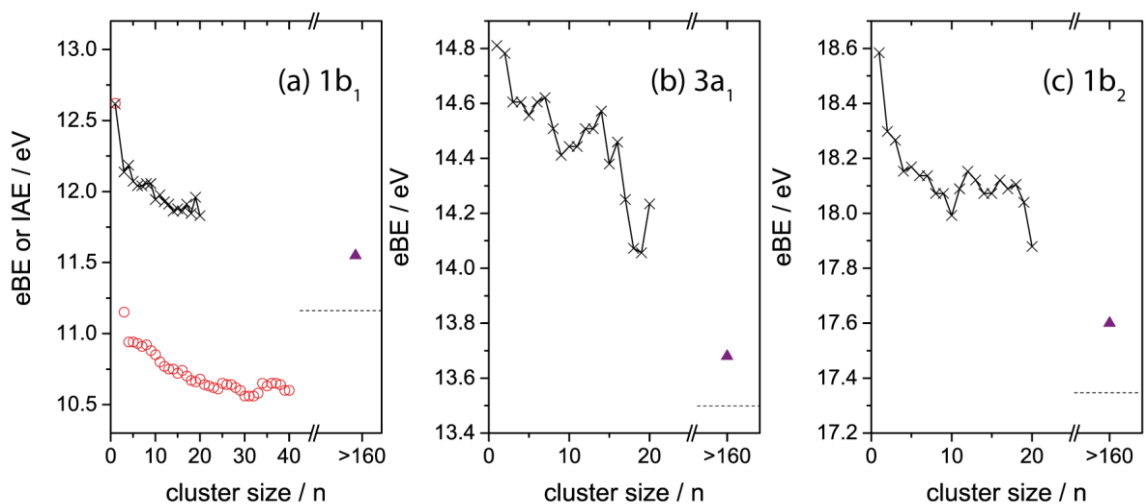

## S2. Scattering calculations for liquid water and clusters

The implementation of the electron scattering model is based on a Monte-Carlo solution of the transport equations [10,11] (see these two refs. for further information). The simulations of detailed angle-resolved photoelectron spectra, which require up to a billion trajectories, were performed with a highly parallel computer program. The modeling of the photoelectron spectra and $\beta$-parameters consists of four main parts which are explained in more detail below: **i)** The interaction of the liquid jet with the ionizing radiation $h\nu$ and the probability of forming quasi-free electrons in the conduction band by ionization of water at each point in the microjet. **ii)** The transport of the electrons from the point of ionization to the liquid jet/cluster surface and the escape from the surface into vacuum. **iii)** The collection of photoelectrons mimicking a typical experimental liquid jet collection geometry or **iv)** The collection of photoelectrons mimicking a typical experimental cluster collection geometry.

**(i)** The probability to generate a quasi-free electron at a certain point in the liquid jet is proportional to the local light intensity of the ionizing radiation $h\nu$. Note that the intensity in small clusters is constant. The electric field inside the liquid jet is calculated from Maxwell's equations for plane-wave irradiation of a cylinder of 5 μm radius and infinite length (liquid microjet, Fig. S5) using the wavelength-dependent complex index of refraction of pure water [12]. (Note that the results are identical for larger jet diameters.) The propagation and polarization direction of the linearly polarized light are shown in Fig. S5. $\theta$ is the angle between the polarization of the ionization laser and the electron detection axis. Maxwell's equations were solved numerically using a finite-difference time-domain (FDTD) code [13].

**Figure S5**: Scheme of a typical liquid jet setup [4,14,15]. The linearly polarized ionization laser (violet arrow) propagates from front to back. $\theta$ is the angle between the laser polarization and the electron detection axis.

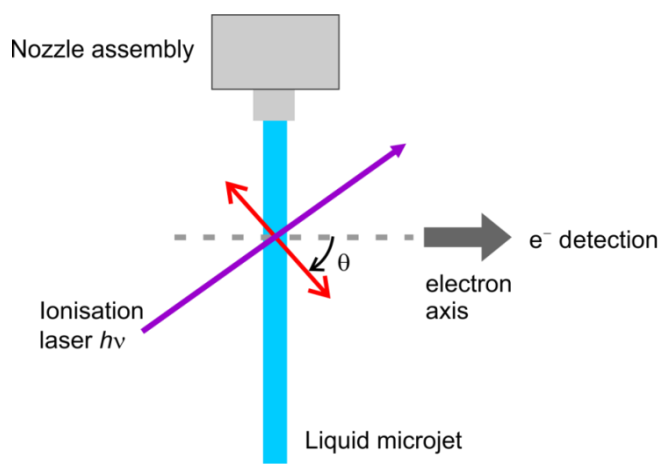

The initial kinetic energy distribution of the quasi-free electrons is described by a sum of Gaussians, whose parameters are chosen so that the calculated spectrum at $h\nu = 60$ eV using the scattering calculations described in (ii) reproduces the experimental photoelectron spectrum of water recorded at $h\nu = 60$ eV by Winter et al. [9].

**(ii)** The probabilistic electron transport model is formulated as a random walk with an exponential distribution of step lengths. The mean step length, i.e. the electron mean free path MFP($E$), depends on the instantaneous kinetic energy $E$ of the electron and is given by $\text{MFP}(E) = \frac{1}{\rho \sigma_{tot}(E)}$, where $\rho$ is the number density of scatterers (water molecules) and $\sigma_{tot}(E)$ is the total scattering cross section. The different scattering events are described by differential scattering cross sections $\sigma(E, \Delta_E, \Omega)$ for energy loss $\Delta_E$ and the deflection angle $\Omega$ of the electron. $\sigma(E, \Delta_E, \Omega)$ is written as the sum of contributions from elastic (i. e. $\Delta_E = 0$) and different types of inelastic scattering (inelastic electron-phonon, electron-vibron, dissociative electron attachment, and electron-electron scattering). The total cross section is given by $\sigma_{tot}(E) = \iint \sigma(E, \Delta_E, \Omega) d\Delta_E d\Omega$. The cross sections used for the present study are taken from ref. [10] for the lower kinetic energy range and from ref. [16] for the higher kinetic energies. The scattering parameters in ref. [10] were determined from photoemission studies of liquid water droplets. The combination of water droplets and velocity map photoelectron imaging (VMI) introduced in ref. [10] allowed us to determine accurate scattering parameters exploiting the detailed information that is contained in the droplet size-dependent photoelectron anisotropies and kinetic energies.[10,17] The scattering parameters in ref. [16] were determined from amorphous ice samples. Note that liquid water scattering data do not yet exist at higher energies, but that there is no reason to expect any difference between amorphous ice and liquid water for the electronic scattering processes in this energy range. For the angular dependence of scattering, we found the representation proposed in ref. [17] for amorphous ice perfectly appropriate also for the liquid [10]. This models the differential cross sections with an explicit angular dependence given by the sum of a θ-independent term ("isotropic contribution") and a |cos(θ)|-term ("forward contribution"). Our analysis of droplet VMIs [10] showed that the effect of elastic scattering is well described by an isotropic contribution alone.

We assume a flat bottom of the conduction band so that electrons within the liquid move at a constant potential below the vacuum level. Consequently, they have to overcome the escape barrier $V_0$ (location of the conduction band edge relative to the vacuum level). The escape barrier is set to $V_0 = 1$ eV, roughly corresponding to the difference between the onsets of photoemission and photoconduction reported for water ice.[16] See also refs. [18,19] for further information on the escape barrier. Electrons with $E < V_0$ are eventually absorbed. The tangential components of the electron's momentum relative to the liquid jet surface is assumed to be conserved when crossing the surface. This leads to a diffraction-like escape condition (Snell's law): The velocity component normal to the liquid jet surface $v_n$ must exceed the escape velocity $v_{esc}$, i.e. $v_{esc} = \sqrt{\dfrac{2V_0}{m_e}}$. Otherwise, the electron is reflected back into the liquid. $m_e$ is the electron mass. Detailed scattering models similar to the present were suggested in refs. [10,16].

**(iii) For the liquid jet calculations**, a typical experimental geometry (Fig. S5) is used for the calculations of the photoelectron kinetic energy distributions (eKEs), the photoelectron binding energy spectra (eBE), and the $\beta$-parameters. eBE spectra are obtained from the eKE distributions from the relation:

$$\text{eBE} = h\nu - \text{eKE} \qquad \text{Eq. (S1)}$$

For the simulations of the angle-resolved photoelectron spectra only the electrons ejected into a small solid angle $\Delta\omega = 9\cdot 10^{-4}$ sr around the electron detection axis $\omega$ are collected. $\Delta\omega$ is determined by the detector's surface area (here 40 mm diameter) and its distance from the point of ionization (here 1200 mm). Angle-resolved photoelectron spectra are simulated by rotating the linear polarization of the ionization laser from $\theta = 0$ to $90°$, where $\theta$ is the angle between the ionization laser polarization and the electron detection axis (Fig. S5). For the description of the photoelectron angular distributions (PADs), we neglect higher terms and use a single anisotropy parameter $\beta$, defined by

$$I(\theta) \propto 1 + \frac{\beta}{2}\left(3\cos^2\theta - 1\right) \qquad \text{Eq. (S2).}$$

Fig. 2 in the main text shows as an example calculated liquid jet spectra for $\theta = 0$ and $90°$ (panel d) together with calculated liquid jet $\beta$-parameters (panel e) for an ionization energy of $h\nu = 38.7$ eV. Fig. 3 in the main text contains the calculated $\beta$-parameters for the liquid as a function of the eKE for the three water orbitals assuming monomer ($\beta_{n=1}$) and cluster ($\beta_{n=6}$) values as inputs for the local values of the anisotropy prameters. We also add a calculation for the liquid in Fig. 3 (open black stars) for which we use the elastic gas phase scattering parameters [20] instead of the proper condensed phase parameters [10,16] as for all other calculations. It deviates strongly from the experimental liquid jet values [4]. Note that below 100 eV, the elastic cross sections in the gas phase are about a factor of 50 larger than in the condensed phase.

For polarization-dependent liquid jet measurements, the $\beta$-parameter is determined from intensity ratios recorded for different $\theta$s; for example from the ratio $\frac{I_{90°}}{I_{0°}}$. This results in artifacts in $\beta$, which arise from the polarization-dependent coupling efficiency of the light into the liquid microjet. For example, at $h\nu \sim 10$ eV the coupling efficiency of the light at 90° is ~ 10% lower than at 0°. Therefore, the determined $\beta$-values are artificially too high by ~ 0.05-0.1 (value for an isotropic input distribution). At $h\nu \sim 30$-40 eV, the coupling efficiency of the light at 90° is ~1-2% lower than at 0°, which reduces the bias in $\beta$ to 0.01-0.02. Furthermore, one should in general keep in mind that the non-spherical symmetry of the liquid jet setup influences the PAD and may lead to deviations compared with PADs measured by other methods, such as velocity map imaging (VMI).

**(iv) For the clusters** (second last paragraph in the main text), the $\beta$-parameters are determined from calculated photoelectron VMIs. The calculations for cluster are equivalent to the calculations for aerosol particles in refs. [10,17] (Eq.(S2)). We performed calculations for n = 1 to $3.5 \cdot 10^6$ at $h\nu$ = 16.7 and 37.4 eV, with the goal to approximately determine at which cluster size elastic and inelastic scattering (contribution (iv) [10,16]) become important in clusters. The calculations show that it needs approximately 100 molecules in a cluster to change $\beta$ by 0.1.

**Figure S6**: Calculated $\beta$-parameters for water clusters with n molecules assuming that only contribution (iv) is present (see main text). The calculations are for two different ionization energies $h\nu$ = 16.7 and 37.4 eV and all three valence orbitals. The $\beta$-values at $h\nu$ = 16.7 are zero because this orbital cannot be ionized at this energy. The used $\beta$-input values are indicated at n = 1.

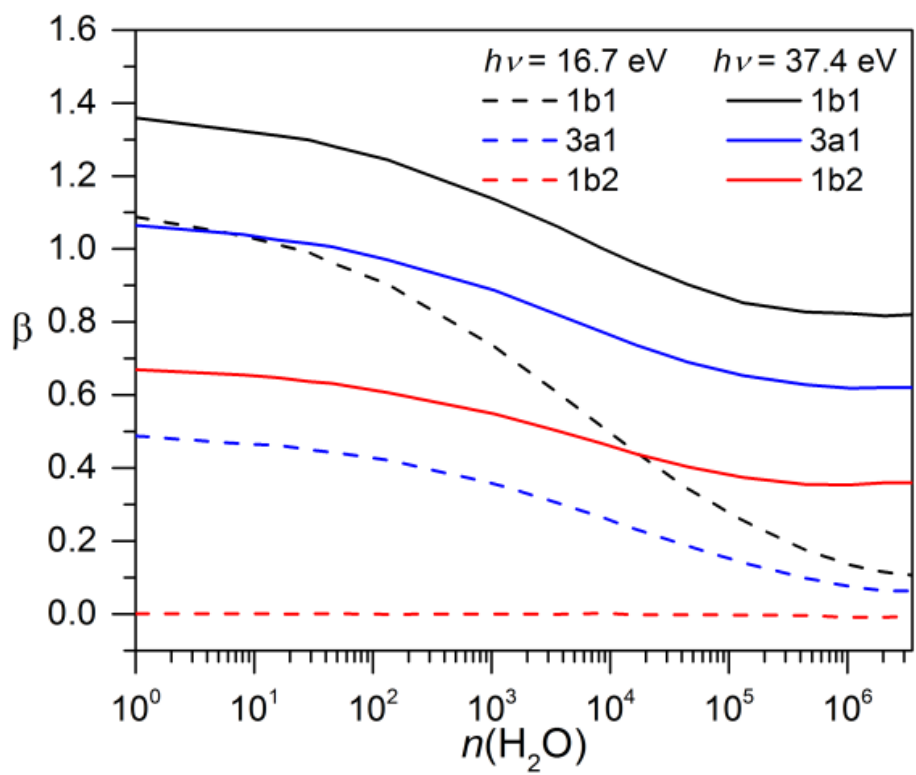

## S3. Additional Tables

**Table T1.**

Published experimental $\beta$-parameters for water for ionization from the valence orbitals $1b_1$, $3a_1$, and $1b_2$ [3,4] and the O1s orbital [15]. This table summarizes gas phase monomer, cluster, and liquid jet data for water. Photon energies $h\nu$ and photoelectron kinetic energies eKEs are given in eV. The average size of the cluster distribution is denoted by <n>. The eKE values for the valence orbitals are determined from the indicated photon energies $h\nu$ and the VBEs reported in ref. [9]. The O1s values are extracted from figure 2 in ref. [15].

|  | $h\nu$ | <n> | $1b_1$ eKE | $1b_1$ $\beta$ | $3a_1$ eKE | $3a_1$ $\beta$ | $1b_2$ eKE | $1b_2$ $\beta$ | O1s eKE | O1s $\beta$ |
|---|---|---|---|---|---|---|---|---|---|---|
| **monomer**[3] | 40 |  | ~27.4 | 1.38(8) | ~25.2 | 1.08(8) | ~21.2 | 0.75(12) |  |  |
| **clusters**[3] |  | 58 |  | 0.83(8) |  | 0.73(16) |  | 0.42(16) |  |  |
| **monomer**[3] | 60 |  | ~47.4 | 1.59(8) | ~45.2 | 1.41(8) | ~41.2 | 1.04(12) |  |  |
| **clusters**[3] |  | >84 |  | 1.17(8) |  | 0.99(12) |  | 0.70(18) |  |  |
| **monomer**[4] | 38.7 |  | ~26.1 | 1.4 | ~23.9 | 1.1 | ~19.8 | 0.7 |  |  |
| **liquid**[4] |  |  | ~27.5 | 0.8 | ~25.2 | 0.7 | ~21.4 | 0.6 |  |  |
| **monomer**[15] |  |  |  |  |  |  |  |  | ~12 | 0.93 |
| **liquid**[15] |  |  |  |  |  |  |  |  | ~12 | 0.28 |
| **monomer**[15] |  |  |  |  |  |  |  |  | ~25 | 1.56 |
| **liquid**[15] |  |  |  |  |  |  |  |  | ~25 | 0.58 |
| **monomer**[15] |  |  |  |  |  |  |  |  | ~55 | 1.83 |
| **liquid**[15] |  |  |  |  |  |  |  |  | ~55 | 1.05 |
| **monomer**[15] |  |  |  |  |  |  |  |  | ~100 | 1.96 |
| **liquid**[15] |  |  |  |  |  |  |  |  | ~100 | 1.38 |
| **monomer**[15] |  |  |  |  |  |  |  |  | ~260 | 2.00 |
| **liquid**[15] |  |  |  |  |  |  |  |  | ~260 | 1.54 |
| **monomer**[15] |  |  |  |  |  |  |  |  | ~460 | 2.00 |
| **liquid**[15] |  |  |  |  |  |  |  |  | ~460 | 1.62 |

**Table T2.**

Experimental $\beta$-parameters for $(H_2O)_n$ water clusters with estimated uncertainties ($\sigma(\beta)$) for ionization from the valence orbitals $1b_1$, $3a_1$, and $1b_2$ from the present work. n is the number of molecules per cluster, $h\nu$ is the photon energy, and eKE is the photoelectron kinetic energy.

| n=1 | | $1b_1$ | | | $3a_1$ | | | $1b_2$ | |
|---|---|---|---|---|---|---|---|---|---|
| $h\nu$ | eKE | $\beta$ | $\sigma(\beta)$ | eKE | $\beta$ | $\sigma(\beta)$ | eKE | $\beta$ | $\sigma(\beta)$ |
| 13.0 | 0.38 | 0.29 | 0.15 | | | | | | |
| 14.0 | 1.38 | 0.63 | 0.11 | | | | | | |
| 15.0 | 2.38 | 0.76 | 0.07 | | | | | | |
| 16.0 | 3.38 | 0.96 | 0.06 | 1.19 | 0.27 | 0.26 | | | |
| 18.0 | 5.38 | 1.15 | 0.03 | 3.19 | 0.51 | 0.18 | | | |
| 20.0 | 7.38 | 1.29 | 0.23 | 5.19 | 0.70 | 0.10 | 1.42 | -0.33 | 0.25 |
| 22.0 | 9.38 | 1.32 | 0.03 | 7.19 | 0.73 | 0.14 | 3.42 | 0.01 | 0.06 |
| 24.0 | 11.38 | 1.32 | 0.24 | 9.19 | 0.81 | 0.14 | 5.42 | 0.24 | 0.10 |
| 27.0 | 14.38 | 1.26 | 0.16 | 12.19 | 0.84 | 0.10 | 8.42 | 0.45 | 0.14 |
| 30.0 | 17.38 | 1.22 | 0.21 | 15.19 | 0.95 | 0.18 | 11.42 | 0.51 | 0.08 |
| 32.5 | 19.88 | 1.50 | 0.21 | 17.69 | 1.14 | 0.17 | 13.92 | 0.61 | 0.15 |
| **n=2** | | $1b_1$ | | | $3a_1$ | | | $1b_2$ | |
| $h\nu$ | eKE | $\beta$ | $\sigma(\beta)$ | eKE | $\beta$ | $\sigma(\beta)$ | eKE | $\beta$ | $\sigma(\beta)$ |
| 13.0 | 0.50 | 0.08 | 0.20 | | | | | | |
| 14.0 | 1.50 | 0.28 | 0.08 | | | | | | |
| 15.0 | 2.50 | 0.52 | 0.08 | | | | | | |
| 16.0 | 3.50 | 0.53 | 0.09 | 1.22 | 0.12 | 0.17 | | | |
| 18.0 | 5.50 | 0.78 | 0.08 | 3.22 | 0.32 | 0.11 | | | |
| 20.0 | 7.50 | 0.86 | 0.09 | 5.22 | 0.54 | 0.13 | 1.70 | -0.18 | 0.18 |
| 22.0 | 9.50 | 0.96 | 0.11 | 7.22 | 0.60 | 0.11 | 3.70 | -0.02 | 0.08 |
| 24.0 | 11.50 | 1.02 | 0.13 | 9.22 | 0.69 | 0.08 | 5.70 | 0.19 | 0.13 |
| 27.0 | 14.50 | 1.16 | 0.22 | 12.22 | 0.68 | 0.17 | 8.70 | 0.51 | 0.20 |
| 30.0 | 17.50 | 1.13 | 0.16 | 15.22 | 0.83 | 0.28 | 11.70 | 0.49 | 0.22 |
| 32.5 | 20.00 | 1.32 | 0.05 | 17.72 | 0.96 | 0.13 | 14.20 | 0.61 | 0.12 |
| **n=3** | | $1b_1$ | | | $3a_1$ | | | $1b_2$ | |
| $h\nu$ | eKE | $\beta$ | $\sigma(\beta)$ | eKE | $\beta$ | $\sigma(\beta)$ | eKE | $\beta$ | $\sigma(\beta)$ |
| 13.0 | 0.86 | 0.10 | 0.20 | | | | | | |
| 14.0 | 1.86 | 0.26 | 0.03 | | | | | | |
| 15.0 | 2.86 | 0.37 | 0.04 | | | | | | |
| 16.0 | 3.86 | 0.39 | 0.05 | 1.40 | 0.06 | 0.23 | | | |
| 18.0 | 5.86 | 0.62 | 0.06 | 3.40 | 0.16 | 0.12 | | | |
| 20.0 | 7.86 | 0.82 | 0.08 | 5.40 | 0.39 | 0.13 | 1.73 | -0.11 | 0.22 |
| 22.0 | 9.86 | 0.97 | 0.08 | 7.40 | 0.46 | 0.10 | 3.73 | 0.03 | 0.10 |
| 24.0 | 11.86 | 0.98 | 0.05 | 9.40 | 0.59 | 0.15 | 5.73 | 0.24 | 0.12 |
| 27.0 | 14.86 | 1.06 | 0.15 | 12.40 | 0.50 | 0.19 | 8.73 | 0.44 | 0.23 |
| 30.0 | 17.86 | 0.97 | 0.14 | 15.40 | 0.69 | 0.27 | 11.73 | 0.46 | 0.14 |
| 32.5 | 20.36 | 1.24 | 0.09 | 17.90 | 0.80 | 0.11 | 14.23 | 0.52 | 0.14 |

| n=4 | | 1b$_1$ | | | 3a$_1$ | | | 1b$_2$ | |
|---|---|---|---|---|---|---|---|---|---|
| hv | eKE | β | σ(β) | eKE | β | σ(β) | eKE | β | σ(β) |
| 13.0 | 0.81 | 0.05 | 0.20 | | | | | | |
| 14.0 | 1.81 | 0.18 | 0.06 | | | | | | |
| 15.0 | 2.81 | 0.24 | 0.07 | | | | | | |
| 16.0 | 3.81 | 0.24 | 0.07 | 1.40 | 0.05 | 0.23 | | | |
| 18.0 | 5.81 | 0.46 | 0.06 | 3.40 | 0.14 | 0.13 | | | |
| 20.0 | 7.81 | 0.67 | 0.07 | 5.40 | 0.35 | 0.12 | 1.85 | -0.10 | 0.21 |
| 22.0 | 9.81 | 0.76 | 0.12 | 7.40 | 0.40 | 0.12 | 3.85 | 0.01 | 0.13 |
| 24.0 | 11.81 | 0.83 | 0.09 | 9.40 | 0.47 | 0.11 | 5.85 | 0.25 | 0.13 |
| 27.0 | 14.81 | 0.97 | 0.26 | 12.40 | 0.41 | 0.34 | 8.85 | 0.41 | 0.24 |
| 30.0 | 17.81 | 0.86 | 0.16 | 15.40 | 0.51 | 0.44 | 11.85 | 0.49 | 0.35 |
| 32.5 | 20.31 | 1.16 | 0.03 | 17.90 | 0.68 | 0.08 | 14.35 | 0.51 | 0.14 |
| **n=5** | | 1b$_1$ | | | 3a$_1$ | | | 1b$_2$ | |
| hv | eKE | β | σ(β) | eKE | β | σ(β) | eKE | β | σ(β) |
| 13.0 | 0.93 | 0.10 | 0.14 | | | | | | |
| 14.0 | 1.93 | 0.11 | 0.06 | | | | | | |
| 15.0 | 2.93 | 0.12 | 0.07 | | | | | | |
| 16.0 | 3.93 | 0.13 | 0.06 | 1.44 | 0.06 | 0.28 | | | |
| 18.0 | 5.93 | 0.33 | 0.05 | 3.44 | 0.13 | 0.18 | | | |
| 20.0 | 7.93 | 0.54 | 0.05 | 5.44 | 0.38 | 0.17 | 1.83 | -0.07 | 0.22 |
| 22.0 | 9.93 | 0.64 | 0.07 | 7.44 | 0.36 | 0.16 | 3.83 | 0.00 | 0.12 |
| 24.0 | 11.93 | 0.66 | 0.08 | 9.44 | 0.52 | 0.16 | 5.83 | 0.21 | 0.17 |
| 27.0 | 14.93 | 0.97 | 0.31 | 12.44 | 0.56 | 0.56 | 8.83 | 0.43 | 0.25 |
| 30.0 | 17.93 | 0.85 | 0.34 | 15.44 | 0.58 | 0.13 | 11.83 | 0.43 | 0.23 |
| 32.5 | 20.43 | 1.07 | 0.04 | 17.94 | 0.73 | 0.23 | 14.33 | 0.58 | 0.16 |
| **n=6** | | 1b$_1$ | | | 3a$_1$ | | | 1b$_2$ | |
| hv | eKE | β | σ(β) | eKE | β | σ(β) | eKE | β | σ(β) |
| 13.0 | 0.96 | 0.06 | 0.16 | | | | | | |
| 14.0 | 1.96 | 0.09 | 0.06 | | | | | | |
| 15.0 | 2.96 | 0.11 | 0.08 | | | | | | |
| 16.0 | 3.96 | 0.11 | 0.09 | 1.40 | 0.04 | 0.29 | | | |
| 18.0 | 5.96 | 0.32 | 0.11 | 3.40 | 0.12 | 0.18 | | | |
| 20.0 | 7.96 | 0.54 | 0.09 | 5.40 | 0.32 | 0.22 | 1.86 | -0.06 | 0.29 |
| 22.0 | 9.96 | 0.68 | 0.07 | 7.40 | 0.36 | 0.35 | 3.86 | 0.01 | 0.17 |
| 24.0 | 11.96 | 0.71 | 0.11 | 9.40 | 0.48 | 0.22 | 5.86 | 0.25 | 0.22 |
| 27.0 | 14.96 | 0.76 | 0.10 | 12.40 | 0.56 | 0.32 | 8.86 | 0.54 | 0.48 |
| 30.0 | 17.96 | 0.77 | 0.10 | 15.40 | 0.57 | 0.29 | 11.86 | 0.49 | 0.14 |
| 32.5 | 20.46 | 1.11 | 0.06 | 17.90 | 0.70 | 0.24 | 14.36 | 0.53 | 0.19 |
| **n=7** | | 1b$_1$ | | | 3a$_1$ | | | 1b$_2$ | |
| hv | eKE | β | σ(β) | eKE | β | σ(β) | eKE | β | σ(β) |
| 13.0 | 0.96 | 0.06 | 0.15 | | | | | | |
| 14.0 | 1.96 | 0.06 | 0.07 | | | | | | |
| 15.0 | 2.96 | 0.10 | 0.10 | | | | | | |
| 16.0 | 3.96 | 0.13 | 0.07 | 1.38 | 0.00 | 0.23 | | | |
| 18.0 | 5.96 | 0.30 | 0.06 | 3.38 | 0.11 | 0.25 | | | |
| 20.0 | 7.96 | 0.49 | 0.08 | 5.38 | 0.29 | 0.28 | 1.86 | -0.05 | 0.25 |

| hv | eKE | β | σ(β) | eKE | β | σ(β) | eKE | β | σ(β) |
|---|---|---|---|---|---|---|---|---|---|
| 22.0 | 9.96 | 0.61 | 0.15 | 7.38 | 0.33 | 0.17 | 3.86 | 0.03 | 0.20 |
| 24.0 | 11.96 | 0.66 | 0.10 | 9.38 | 0.33 | 0.15 | 5.86 | 0.23 | 0.20 |
| 27.0 | 14.96 | 0.98 | 0.15 | 12.38 | 0.66 | 0.81 | 8.86 | 0.41 | 0.28 |
| 30.0 | 17.96 | 0.83 | 0.13 | 15.38 | 0.50 | 0.34 | 11.86 | 0.40 | 0.44 |
| 32.5 | 20.46 | 1.00 | 0.04 | 17.88 | 0.68 | 0.21 | 14.36 | 0.46 | 0.15 |
| **n=8** | | $1b_1$ | | | $3a_1$ | | | $1b_2$ | |
| hv | eKE | β | σ(β) | eKE | β | σ(β) | eKE | β | σ(β) |
| 13.0 | 0.94 | 0.06 | 0.15 | | | | | | |
| 14.0 | 1.94 | 0.09 | 0.10 | | | | | | |
| 15.0 | 2.94 | 0.09 | 0.10 | | | | | | |
| 16.0 | 3.94 | 0.12 | 0.07 | 1.49 | -0.02 | 0.28 | | | |
| 18.0 | 5.94 | 0.31 | 0.10 | 3.49 | 0.12 | 0.27 | | | |
| 20.0 | 7.94 | 0.48 | 0.09 | 5.49 | 0.30 | 0.31 | 1.93 | -0.05 | 0.31 |
| 22.0 | 9.94 | 0.59 | 0.06 | 7.49 | 0.27 | 0.20 | 3.93 | 0.00 | 0.19 |
| 24.0 | 11.94 | 0.65 | 0.08 | 9.49 | 0.38 | 0.16 | 5.93 | 0.24 | 0.20 |
| 27.0 | 14.94 | 0.97 | 0.15 | 12.49 | 0.53 | 0.44 | 8.93 | 0.47 | 0.60 |
| 30.0 | 17.94 | 0.83 | 0.23 | 15.49 | 0.49 | 0.31 | 11.93 | 0.39 | 0.18 |
| 32.5 | 20.44 | 0.99 | 0.02 | 17.99 | 0.63 | 0.10 | 14.43 | 0.47 | 0.05 |
| **n=9** | | $1b_1$ | | | $3a_1$ | | | $1b_2$ | |
| hv | eKE | β | σ(β) | eKE | β | σ(β) | eKE | β | σ(β) |
| 13.0 | 0.94 | 0.04 | 0.17 | | | | | | |
| 14.0 | 1.94 | 0.05 | 0.10 | | | | | | |
| 15.0 | 2.94 | 0.06 | 0.12 | | | | | | |
| 16.0 | 3.94 | 0.09 | 0.08 | 1.59 | 0.00 | 0.31 | | | |
| 18.0 | 5.94 | 0.29 | 0.06 | 3.59 | 0.16 | 0.22 | | | |
| 20.0 | 7.94 | 0.51 | 0.12 | 5.59 | 0.31 | 0.38 | 1.93 | 0.02 | 0.44 |
| 22.0 | 9.94 | 0.59 | 0.10 | 7.59 | 0.27 | 0.37 | 3.93 | 0.04 | 0.34 |
| 24.0 | 11.94 | 0.63 | 0.07 | 9.59 | 0.38 | 0.27 | 5.93 | 0.21 | 0.22 |
| 27.0 | 14.94 | 0.75 | 0.19 | 12.59 | 0.44 | 0.28 | 8.93 | 0.26 | 0.44 |
| 30.0 | 17.94 | 0.77 | 0.14 | 15.59 | 0.41 | 0.17 | 11.93 | 0.35 | 0.13 |
| 32.5 | 20.44 | 1.07 | 0.08 | 18.09 | 0.73 | 0.24 | 14.43 | 0.57 | 0.21 |
| **n=10** | | $1b_1$ | | | $3a_1$ | | | $1b_2$ | |
| hv | eKE | β | σ(β) | eKE | β | σ(β) | eKE | β | σ(β) |
| 13.0 | 1.06 | 0.03 | 0.18 | | | | | | |
| 14.0 | 2.06 | 0.05 | 0.11 | | | | | | |
| 15.0 | 3.06 | 0.08 | 0.10 | | | | | | |
| 16.0 | 4.06 | 0.12 | 0.07 | 1.56 | -0.01 | 0.29 | | | |
| 18.0 | 6.06 | 0.25 | 0.10 | 3.56 | 0.09 | 0.26 | | | |
| 20.0 | 8.06 | 0.54 | 0.11 | 5.56 | 0.32 | 0.32 | 2.01 | 0.00 | 0.34 |
| 22.0 | 10.06 | 0.61 | 0.15 | 7.56 | 0.33 | 0.26 | 4.01 | -0.02 | 0.19 |
| 24.0 | 12.06 | 0.62 | 0.03 | 9.56 | 0.52 | 0.11 | 6.01 | 0.16 | 0.13 |
| 27.0 | 15.06 | 0.82 | 0.23 | 12.56 | 0.45 | 0.32 | 9.01 | 0.52 | 0.28 |
| 30.0 | 18.06 | 0.79 | 0.10 | 15.56 | 0.62 | 0.23 | 12.01 | 0.35 | 0.43 |
| 32.5 | 20.56 | 0.99 | 0.07 | 18.06 | 0.64 | 0.23 | 14.51 | 0.54 | 0.16 |

| n=11 | | 1b$_1$ | | | 3a$_1$ | | | 1b$_2$ | |
|---|---|---|---|---|---|---|---|---|---|
| hv | eKE | β | σ(β) | eKE | β | σ(β) | eKE | β | σ(β) |
| 13.0 | 1.02 | 0.01 | 0.15 | | | | | | |
| 14.0 | 2.02 | 0.06 | 0.16 | | | | | | |
| 15.0 | 3.02 | 0.06 | 0.09 | | | | | | |
| 16.0 | 4.02 | 0.07 | 0.10 | 1.56 | -0.05 | 0.30 | | | |
| 18.0 | 6.02 | 0.28 | 0.09 | 3.56 | 0.11 | 0.30 | | | |
| 20.0 | 8.02 | 0.45 | 0.10 | 5.56 | 0.25 | 0.27 | 1.91 | -0.06 | 0.30 |
| 22.0 | 10.02 | 0.62 | 0.15 | 7.56 | 0.28 | 0.27 | 3.91 | 0.03 | 0.38 |
| 24.0 | 12.02 | 0.61 | 0.07 | 9.56 | 0.51 | 0.28 | 5.91 | 0.18 | 0.29 |
| 27.0 | 15.02 | 1.01 | 0.29 | 12.56 | 0.54 | 0.78 | 8.91 | 0.45 | 0.42 |
| 30.0 | 18.02 | 0.91 | 0.03 | 15.56 | 0.46 | 0.57 | 11.91 | 0.41 | 0.25 |
| 32.5 | 20.52 | 1.01 | 0.09 | 18.06 | 0.77 | 0.46 | 14.41 | 0.47 | 0.19 |
| n=12 | | 1b$_1$ | | | 3a$_1$ | | | 1b$_2$ | |
| hv | eKE | β | σ(β) | eKE | β | σ(β) | eKE | β | σ(β) |
| 13.0 | 1.07 | -0.01 | 0.13 | | | | | | |
| 14.0 | 2.07 | 0.06 | 0.10 | | | | | | |
| 15.0 | 3.07 | 0.05 | 0.07 | | | | | | |
| 16.0 | 4.07 | 0.11 | 0.17 | 1.49 | -0.03 | 0.27 | | | |
| 18.0 | 6.07 | 0.25 | 0.10 | 3.49 | 0.07 | 0.24 | | | |
| 20.0 | 8.07 | 0.43 | 0.11 | 5.49 | 0.23 | 0.32 | 1.85 | -0.09 | 0.44 |
| 22.0 | 10.07 | 0.63 | 0.30 | 7.49 | 0.34 | 0.38 | 3.85 | 0.01 | 0.30 |
| 24.0 | 12.07 | 0.61 | 0.08 | 9.49 | 0.45 | 0.25 | 5.85 | 0.20 | 0.22 |
| n=13 | | 1b$_1$ | | | 3a$_1$ | | | 1b$_2$ | |
| hv | eKE | β | σ(β) | eKE | β | σ(β) | eKE | β | σ(β) |
| 13.0 | 1.09 | 0.01 | 0.15 | | | | | | |
| 14.0 | 2.09 | 0.04 | 0.11 | | | | | | |
| 15.0 | 3.09 | 0.03 | 0.09 | | | | | | |
| 16.0 | 4.09 | 0.10 | 0.08 | 1.49 | -0.03 | 0.34 | | | |
| 18.0 | 6.09 | 0.25 | 0.07 | 3.49 | 0.09 | 0.26 | | | |
| 20.0 | 8.09 | 0.46 | 0.11 | 5.49 | 0.19 | 0.32 | 1.88 | 0.00 | 0.57 |
| 22.0 | 10.09 | 0.62 | 0.13 | 7.49 | 0.27 | 0.30 | 3.88 | 0.05 | 0.42 |
| 24.0 | 12.09 | 0.61 | 0.02 | 9.49 | 0.33 | 0.40 | 5.88 | 0.14 | 0.27 |
| n=14 | | 1b$_1$ | | | 3a$_1$ | | | 1b$_2$ | |
| hv | eKE | β | σ(β) | eKE | β | σ(β) | eKE | β | σ(β) |
| 13.0 | 1.14 | -0.01 | 0.12 | | | | | | |
| 14.0 | 2.14 | 0.09 | 0.14 | | | | | | |
| 15.0 | 3.14 | 0.07 | 0.12 | | | | | | |
| 16.0 | 4.14 | 0.10 | 0.09 | 1.43 | -0.04 | 0.32 | | | |
| 18.0 | 6.14 | 0.22 | 0.13 | 3.43 | 0.16 | 0.69 | | | |
| 20.0 | 8.14 | 0.47 | 0.21 | 5.43 | 0.21 | 0.36 | 1.93 | -0.07 | 0.46 |
| 22.0 | 10.14 | 0.60 | 0.12 | 7.43 | 0.30 | 0.47 | 3.93 | 0.03 | 0.38 |
| 24.0 | 12.14 | 0.64 | 0.04 | 9.43 | 0.53 | 0.28 | 5.93 | 0.16 | 0.25 |

| n=15 | | 1b$_1$ | | | 3a$_1$ | | | 1b$_2$ | |
|---|---|---|---|---|---|---|---|---|---|
| hν | eKE | β | σ(β) | eKE | β | σ(β) | eKE | β | σ(β) |
| 13.0 | 1.12 | -0.01 | 0.12 | | | | | | |
| 14.0 | 2.12 | 0.07 | 0.18 | | | | | | |
| 15.0 | 3.12 | 0.06 | 0.09 | | | | | | |
| 16.0 | 4.12 | 0.06 | 0.12 | 1.62 | -0.04 | 0.31 | | | |
| 18.0 | 6.12 | 0.28 | 0.35 | 3.62 | 0.10 | 0.60 | | | |
| 20.0 | 8.12 | 0.49 | 0.18 | 5.62 | 0.31 | 0.45 | 1.93 | -0.10 | 0.36 |
| 22.0 | 10.12 | 0.61 | 0.28 | 7.62 | 0.26 | 0.36 | 3.93 | 0.04 | 0.39 |
| 24.0 | 12.12 | 0.71 | 0.22 | 9.62 | 0.48 | 0.36 | 5.93 | 0.25 | 0.39 |
| **n=16** | | 1b$_1$ | | | 3a$_1$ | | | 1b$_2$ | |
| hν | eKE | β | σ(β) | eKE | β | σ(β) | eKE | β | σ(β) |
| 13.0 | 1.20 | 0.01 | 0.14 | | | | | | |
| 14.0 | 2.20 | 0.06 | 0.16 | | | | | | |
| 15.0 | 3.20 | 0.04 | 0.09 | | | | | | |
| 16.0 | 4.20 | 0.05 | 0.06 | 1.70 | 0.05 | 0.22 | | | |
| 18.0 | 6.20 | 0.24 | 0.13 | 3.70 | 0.00 | 0.25 | | | |
| 20.0 | 8.20 | 0.39 | 0.13 | 5.70 | 0.09 | 0.31 | 2.00 | -0.21 | 0.68 |
| 22.0 | 10.20 | 0.56 | 0.14 | 7.70 | 0.35 | 0.55 | 4.00 | -0.01 | 0.20 |
| 24.0 | 12.20 | 0.62 | 0.09 | 9.70 | 0.56 | 0.27 | 6.00 | 0.12 | 0.19 |
| **n=17** | | 1b$_1$ | | | 3a$_1$ | | | 1b$_2$ | |
| hν | eKE | β | σ(β) | eKE | β | σ(β) | eKE | β | σ(β) |
| 13.0 | 1.09 | 0.02 | 0.16 | | | | | | |
| 14.0 | 2.09 | 0.06 | 0.10 | | | | | | |
| 15.0 | 3.09 | 0.01 | 0.09 | 0.75 | 0.22 | 0.43 | | | |
| 16.0 | 4.09 | 0.08 | 0.07 | 1.75 | 0.05 | 0.31 | | | |
| 18.0 | 6.09 | 0.22 | 0.08 | 3.75 | 0.02 | 0.16 | | | |
| 20.0 | 8.09 | 0.44 | 0.09 | 5.75 | 0.19 | 0.16 | 1.91 | -0.08 | 0.33 |
| 22.0 | 10.09 | 0.56 | 0.06 | 7.75 | 0.33 | 0.13 | 3.91 | 0.06 | 0.30 |
| 24.0 | 12.09 | 0.67 | 0.04 | 9.75 | 0.29 | 0.13 | 5.91 | 0.31 | 0.21 |
| **n=18** | | 1b$_1$ | | | 3a$_1$ | | | 1b$_2$ | |
| hν | eKE | β | σ(β) | eKE | β | σ(β) | eKE | β | σ(β) |
| 13.0 | 1.15 | 0.01 | 0.12 | | | | | | |
| 14.0 | 2.15 | 0.03 | 0.16 | | | | | | |
| 15.0 | 3.15 | 0.00 | 0.09 | 0.93 | 0.12 | 0.42 | | | |
| 16.0 | 4.15 | 0.07 | 0.11 | 1.93 | 0.09 | 0.23 | | | |
| 18.0 | 6.15 | 0.24 | 0.08 | 3.93 | 0.05 | 0.15 | | | |
| 20.0 | 8.15 | 0.45 | 0.04 | 5.93 | 0.14 | 0.20 | 1.90 | -0.07 | 0.42 |
| 22.0 | 10.15 | 0.50 | 0.11 | 7.93 | 0.31 | 0.34 | 3.90 | 0.09 | 0.43 |
| 24.0 | 12.15 | 0.58 | 0.09 | 9.93 | 0.37 | 0.15 | 5.90 | 0.19 | 0.19 |
| **n=19** | | 1b$_1$ | | | 3a$_1$ | | | 1b$_2$ | |
| hν | eKE | β | σ(β) | eKE | β | σ(β) | eKE | β | σ(β) |
| 13.0 | 1.04 | 0.02 | 0.21 | | | | | | |
| 14.0 | 2.04 | 0.10 | 0.12 | | | | | | |
| 15.0 | 3.04 | -0.05 | 0.07 | 0.94 | 0.12 | 0.43 | | | |
| 16.0 | 4.04 | 0.08 | 0.11 | 1.94 | 0.17 | 0.34 | | | |
| 18.0 | 6.04 | 0.25 | 0.11 | 3.94 | 0.05 | 0.21 | | | |
| 20.0 | 8.04 | 0.41 | 0.04 | 5.94 | 0.16 | 0.29 | 1.96 | -0.06 | 0.30 |
| 22.0 | 10.04 | 0.52 | 0.15 | 7.94 | 0.22 | 0.18 | 3.96 | 0.08 | 0.28 |
| 24.0 | 12.04 | 0.62 | 0.11 | 9.94 | 0.37 | 0.16 | 5.96 | 0.31 | 0.45 |
| **n=20** | | $1b_1$ | | | $3a_1$ | | | $1b_2$ | |
| $h\nu$ | eKE | $\beta$ | $\sigma(\beta)$ | eKE | $\beta$ | $\sigma(\beta)$ | eKE | $\beta$ | $\sigma(\beta)$ |
| 13.0 | 1.17 | -0.01 | 0.16 | | | | | | |
| 14.0 | 2.17 | 0.03 | 0.14 | | | | | | |
| 15.0 | 3.17 | 0.04 | 0.12 | 0.77 | -0.10 | 0.52 | | | |
| 16.0 | 4.17 | 0.07 | 0.11 | 1.77 | 0.02 | 0.34 | | | |
| 18.0 | 6.17 | 0.21 | 0.18 | 3.77 | 0.07 | 0.54 | | | |
| 20.0 | 8.17 | 0.45 | 0.09 | 5.77 | 0.22 | 0.21 | 2.12 | -0.13 | 0.23 |
| 22.0 | 10.17 | 0.54 | 0.13 | 7.77 | 0.42 | 0.20 | 4.12 | -0.01 | 0.21 |
| 24.0 | 12.17 | 0.47 | 0.14 | 9.77 | 0.43 | 0.47 | 6.12 | 0.17 | 0.40 |

**Table T3.**

The table summarizes values from the present work and literature values for vertical binding energy (VBE) from photoelectron spectroscopy, adiabatic ionization energy (AIE) from photoelectron spectroscopy, and ion appearance energy (IAE) from ion mass spectrometry for the water dimer. The ionization of $(H_2O)_2$ leads to the formation of $(H_2O)_2^+$ and $H_3O^+$ (fast proton transfer, Eq. (2) main text). For the dimer, ionization from the lone pair of the hydrogen-bond donor $(b_1)_D$ is distinguishable from ionization of the mixed $(a_1/b_1)$ orbital, which is delocalized over donor and acceptor (see main text).

| | $(H_2O)_2^+$ / $(b_1)_D$ | $H_3O^+$ / $(a_1/b_1)$ | reference |
|---|---|---|---|
| VBE | 11.66(10) | 12.98(10) | this work |
| | 12.1 ± 0.1 | 13.2 ± 0.2 | [21] |
| | 11.72 | 13.16 | [22] |
| AIE | 11.15(10) | | this work |
| | 11.1 | | [21] |
| | 11.1 | 12.5 | [22] |
| IAE | 11.21 ± 0.09 | 11.73 ± 0.03 | [23] |
| | 11.25 ± 0.05 | 11.74 ± 0.05 | [7] |
| | | 11.756 ± 0.002 | [24] |